\documentclass[prd, aps, superscriptaddress, preprintnumbers, twocolumn, floatfix, nofootinbib]{revtex4}
\pdfoutput=1

\usepackage{amsfonts}
\usepackage{amsmath}
\usepackage{amssymb}
\usepackage{bm}
\usepackage{dcolumn}
\usepackage{graphicx}   
\usepackage[latin1]{inputenc}
\usepackage{latexsym}
\usepackage{rotating}
\usepackage{hyperref}
\usepackage{graphicx}
\usepackage{color}
\usepackage{tensor}
\usepackage{braket}

\newcommand\be{\begin{equation}}
\newcommand\ba{\begin{eqnarray}}
\newcommand\ee{\end{equation}}
\newcommand\ea{\end{eqnarray}}

\begin{document}

\title{Note on Shape Moduli Stabilization, String Gas Cosmology and the Swampland Criteria}

\author{Gabrielle A. Mitchell}
\email{gabrielle.mitchell@mail.mcgill.ca}
\affiliation{Department of Physics, McGill University, Montr\'{e}al, QC, H3A 2T8, Canada}

\author{Robert Brandenberger}
\email{rhb@physics.mcgill.ca}
\affiliation{Department of Physics, McGill University, Montr\'{e}al, QC, H3A 2T8, Canada}

\date{\today}

\begin{abstract}
 
In String Gas Cosmology, the simplest shape modulus fields are naturally stabilized by taking into account the  presence of string winding and momentum modes. We determine the resulting effective potential for these fields and show that it obeys the {\it de Sitter} conjecture, one of the {\it swampland criteria} for effective field theories to be consistent with superstring theory.
 
\end{abstract}

\maketitle


\section{Introduction}

In recent years there has been a lot of interest in constraints on low energy effective field theories which can emerge from superstring theory. These constraints pick out a small subspace (the {\it string landscape}) of the huge space of possible effective field theories. Effective field theories which are not embeddable in string theory are said to lie in the {\it swampland} (see e.g. \cite{Brennan, Palti} for reviews). Two key criteria are the {\it distance conjecture} and the {\it de Sitter} constraint. The {\it distance conjecture} \cite{Vafa1} states that an effective field theory of a canonically normalized scalar field $\phi$ is only consistent with string theory if the field range $\Delta \phi$ is smaller than $c_1 m_{pl}$, where $m_{pl}$ is the four space-time dimensional Planck mass, and $c_1$ is a constant of order one. The {\it de Sitter condition} \cite{Vafa2} states that the potential $V(\phi)$ of such a field has to be sufficiently steep, i.e.
\be \label{cond2a}
|\frac{V^{\prime}}{V}| \, < \, \frac{c_2}{m_{pl}} \, ,
\ee
where $c_2$ is another constant of the order one, and a prime denotes the derivative with respect to the field. In the case that the potential has a local extremum, then the condition (\ref{cond2a}) is may not be met, but in that case an extended version of the criterion applies which states that \cite{Krishnan, Vafa3}
\be \label{cond2b}
\frac{V^{\prime \prime}}{V} \, < \, - \frac{c_3}{m_{pl}^2} \, ,
\ee
where $c_3$ is a positive constant of the order one.

The swampland conjectures have been formulated based on well-motivated ideas from string theory, but they have not yet been rigorously established. Assuming string theory, all scalar fields appearing in a low energy effective action are {\it moduli fields} of the string compactification, e.g. the radii and shape parameters of extra dimensions. Hence, their potentials are determined by string theory. In a recent paper \cite{Samuel} the effective potential of the scalar field corresponding to the radius of an extra dimension was studied. The starting point was the {\it String Gas Cosmology} model \cite{BV} (see also \cite{Perlt} for earlier work, and \cite{SGCrevs} for reviews), in which matter is described by a gas of strings including both momentum and winding modes, and is coupled to a background space-time. It is known \cite{Patil} that the radion modulus is stabilized by the presence of both winding and momentum modes. The momentum modes prevent the radion from decreasing to zero while the winding modes prevent it from expanding without limits. The resulting effective potential for the radion has vanishing potential energy at its minimum. We found that the effective potential is quadratic about the minimum and hence satisfies the de Sitter criteria (\ref{cond2a}, \ref{cond2b}). 

In this paper, we will focus on the shape moduli. We follow \cite{Edna} and consider two internal toroidal dimensions with radius $R$ and angle $\theta$. From the work of \cite{Edna} it is known that string effects stabilize the shape modulus field. Here we will consider the effective potential for $\theta$ and show that it is also consistent with the conditions (\ref{cond2a}, \ref{cond2b}). 
 
In the following section we give a brief review of SGC. In Section III we derive the effective potential for our shape modulus field $\theta$ and show that it obeys the swampland criteria. We conclude with a discussion of our results. We work in natural units in which the speed of light and Planck's constant are set to $1$. We also set the string length equal to $1$ in our units. 

\section{Brief Review of String Gas Cosmology}
  
String gas cosmology is based upon coupling a classical background (including the graviton and dilaton fields) to a gas of strings. Strings have three types of states: momentum modes, oscillatory modes, and winding modes. These string states, as well as the T-duality symmetry, are the key features of string theory that are used to develop string gas cosmology. String theory requires six extra spatial dimensions, which we take to be compactified and toroidal. Namely, the space-time metric is:
\begin{equation}
\label{eqn:met}
ds^2 = \tensor*{g}{*_{\mu \nu}}\tensor*{dx}{*^{\mu}}\tensor*{dx}{*^{\nu}} + \tensor*{\gamma}{*_{a b}}\tensor*{dx}{*^{a}}\tensor*{dx}{*^{b}}
\end{equation}
where the Latin indices label the compactified dimensions, and the Greek indices label the four dimensional FRW metric. The torus is parametrized by shape and size moduli.

A major challenge in string cosmology is stabilizing the string moduli, that is, stabilizing the sizes and shapes of the extra dimensions, as well as the dilaton. Note that we are here considering moduli stabilization at late times when the background can be described by dilaton gravity \footnote{In the initial high temperature Hagedorn phase of strings, dilaton gravity is inapplicable since it is not consistent with the basic symmetries of string theory. For a recent attempt to construct a background which is consistent with the T-duality symmetry of string theory see e.g. \cite{Guilherme}.}. The basic principle of size modulus stabilization is that the winding modes prevent expansion because their energies increase with $R$, whereas the momentum modes prevent contraction because their energies increase with $\frac{1}{R}$ \cite{Patil}. As we will review here, the coupling of the string gas to the background also provides a stabilization mechanism for a simple shape modulus field \cite{Edna}. It can be shown \cite{Frey} that nonperturbative effects due to gaugino condensation can stabilize the dilaton without interfering with size and shape modulus stabilization, and the same mechanism leads to high scale supersymmetry breaking \cite{Wei}. 

\section{Shape Modulus Potential}

In the following we consider our space-time dimensions to be non-compact, and add a number of compact dimensions which we take to be toroidal. Strings have momentum and winding numbers about the extra dimensions.

The matter action for a mas of strings at temperature $\beta^{-1}$ in D space-time dimensions is given by
\be \label{action1}
S \, = \, \int d^D x \sqrt{-G} \sum_{n^a, w_a, N, {\tilde N}, {\bf{p}}_{nc}} 
\mu_{n^a, w_a, N, {\tilde N}, {\bf{p}}_{nc}} \epsilon_{n^a, w_a, N, {\tilde N}, {\bf{p}}_{nc}} \, ,
\ee
where $G$ is the determinant of the full metric, and the sum runs over all free string states. These states are labelled by their momentum numbers $n^a$, winding numbers $w_a$, numbers $N$ and $\tilde{N}$ of right- and left-moving oscillatory modes, and momentum ${\cal{p}}_{nc}$ in the non-compact directions. The quantity $\mu_{n^a, w_a, N, {\tilde N}, {\bf{p}}_{nc}}$ is the number density of a state with the specified quantum numbers, and $\epsilon_{n^a, w_a, N, {\tilde N}, {\bf{p}}_{nc}}$ is its energy, given by
\be
\epsilon_{n^a, w_a, N, {\tilde N}, {\bf{p}}_{nc}} \, = \, \sqrt{{\bf{p}}_{nc}^2  + M^2_{n^a, w_a, N, {\tilde N}}} \, ,
\ee
where $M_{n^a, w_a, N, {\tilde N}}$ is the mass of such a state. In the absence of string interactions, the number density of states is constant in comoving coordinates. We denote the comoving number density by $n_{n^a, w_a, N, {\tilde N}}$. In thermal equilibrium, the number density of excited states is suppressed by the thermal Boltzmann factor. Hence, we can restrict the summation to run over the lowest mass states which are the massless states (the tachyon which appears in bosonic string theory does not appear in the spectrum of superstring theory). Since the determimant of the metric reduced to our expanding spatial dimensions cancels out between $\sqrt{G}$ and $\mu$, the matter action becomes
\be \label{action2}
S \, = \, \int d^D x \sqrt{- G_{00}} \sum_{{\rm{restricted}}}
n_{n^a, w_a, N, {\tilde N}, {\bf{p}}_{nc}} \epsilon_{n^a, w_a, N, {\tilde N}, {\bf{p}}_{nc}} \,
\ee
where the sum is restricted to the lowest mass states, and $G_{00}$ is the time component of the metric (the scale factor dependence in the spatial dimensions has been cancelled against the corresponding factor appearing in the number densities $\mu$).

In \cite{Patil} and \cite{Edna}, the equations of motion for the size and shape moduli were derived by deriving the energy-momentum tensor of the higher dimensional theory with matter action (\ref{action2}), and inserting the resulting expressions into the higher dimensional Einstein action. Here, following \cite{Samuel}, we obtain the equations of motion for the moduli fields by inserting the ansatz for the metric into the matter action (\ref{action2}) (more specifically, by inserting the ansatz into the mass formula), viewing the resulting action as an action for the moduli fields, and taking the resulting variational equations.

The string mass for a toroidal compactification depends on the three sets of string quantum numbers, and is given by:
\ba
\label{eqn:mass}
M^2_{\vec{n}, \vec{n}, \vec{N}} \, &=& \, \frac{1}{R^2}\gamma ^{ab}n_an_b + \frac{R^2}{\alpha '^2}\gamma _{ab}w^aw^b \nonumber \\
&& \, + \frac{2}{\alpha '}\left( 2N + n^aw_a -2 \right) \, ,
\ea
where $n^a$, $w^a$, and $N$ are the momentum, winding, and oscillatory quantum numbers, respectively \footnote{More precisely, $N$ is the number of right-moving modes, and we have made use of the level matching condition to take into account the left-moving modes.}, and $\gamma_{a b}$ is the space-time metric. In the string units which we are using, $\alpha^{\prime} = 1$. The indices $a$ and $b$ run over all compact dimensions. Later in this note we will be considering a two dimensional torus representing two of the internal dimensions. 

In SGC, the universe begins in a thermal state of strings \footnote{Thermal fluctuations of strings in the early phase provide the origin for the cosmological fluctuations \cite{NBV} and gravitational waves \cite{BNPV} observed today.} The string partition function is then dominated by the string states with lowest mass. These states satisfy $N=1$ and $n^a = w_a = \pm 1 $. Note that for each such state, there are momenta and windings in only one of the directions.

If matter is treated as a gas of strings, then the dynamics is governed by the following $d$ space-time dimensional full low energy effective action (the action for matter and geometry):
\ba
\label{eqn:act}
&& S  \, = \,  \frac{1}{2 \kappa _0 ^2} \int d^dX \sqrt{-G} e^{-2 \Phi _d} \\
&& \left[ \hat{R}^d + 4 \partial _\mu \Phi _d \partial ^\mu \Phi _d  
 \, -\frac{1}{4}\partial _\mu \gamma _{ac} \partial ^\mu \gamma ^{ab} - 2 \kappa _0 ^2 e^{-2 \Phi _d}n \braket {E_1} \right] \, , \nonumber
\ea
where $\kappa _0^{-1}$ is the reduced gravitational constant, $n$ is the number density of the strings, $G$ is the determinant of the metric, $\Phi$ is the dimensionless dilaton field, $\hat{R}$ is the Ricci scalar, and $\braket {E_1}$ is the thermal average of the energy of a single string. We consider this as the action for the moduli fields.

The final term in (\ref{eqn:act}) contains no derivatives of the moduli fields and hence acts as a potential for these fields, i.e.
\begin{equation}
V(\phi) \, = \, e^{2 \Phi _d}n \braket{E(\phi)}  \, .
\end{equation}
Since the partition function is dominated by the lowest mass string states, the potential energy can be expressed in terms of the mass of these states, i.e.
\begin{equation}
\label{eqn:potential}
V(\phi) \, \sim \,  e^{2 \Phi _d}n \sqrt{{\bf{p}}_{nc}^2 + M_{1, -1, 1}^2} \, .
\end{equation}
We will  now review how this potential stabilizes the moduli fields, and extract the shape of the potential for a canonical shape modulus field.
 
\section{Moduli stabilization on a 2-dimensional torus}

To be specific, we shall here consider the case of an internal two dimensional torus. Thus, the metric we will be using is that of equation (\ref{eqn:met}) where $\gamma _{ab}$ is the metric for the torus:
\begin{equation}
\label{eqn:metric}
\gamma_{ab} = \begin{bmatrix}
R^2 & R^2sin\theta \\
R^2sin\theta & R^2 \, ,
\end{bmatrix}
\end{equation} 
where $R$ gives the radius of the torus, and $\theta$ is the shape parameter. For $\theta = 0$, it is easy to see that the square mass function (\ref{eqn:mass}) has a minimum at the self-dual radius $R = 1$, and hence leads to stablization of the radial modulus field $R$, as has been considered in \cite{Patil}. The resulting modulus potential was shown \cite{Samuel} to be consistent with the de Sitter conjecture. Here, we focus on the shape modulus and its potential.

The metric $\gamma_{ab}$ generally involves scalar fields $\phi ^I$ called moduli fields. The kinetic terms of the moduli fields are:
\begin{equation}
\label{eqn:mod}
- \frac{1}{4} \partial _ \mu \gamma_{ac} \partial ^ \mu \gamma ^{ab} = -g_{IJ} \partial _ \mu \phi ^{I} \partial _ \mu \phi ^{J}
\end{equation}
In our case, we fix $R$ and hence consider only one moduli field. It corresponds to the shape modulus $\theta$. The corresponding canonically normalized field is 
\be \label{range}
 \phi \, \equiv \, \frac{M_{pl}}{R} \theta \, ,
 \ee
 which leads to the internal space metric
\begin{equation}
g_{IJ} \, = \, \begin{bmatrix}
\frac{1}{4} & 0\\
0 & \frac{1}{4}
\end{bmatrix}
\end{equation}
in the limit of small $\phi$. Note that $\phi = 0$ corresponds to a rectangular torus, the enhanced symmetry point.

The stable fixed point is the rectangular torus where the complex structure modulus $\theta$ is zero, and $R$ is set to unity. We can explicitly find the string mass for the metric $\gamma _{ab}$ defined in (\ref{eqn:metric}). Expanding equation (\ref{eqn:mass}) about small field values gives:
\begin{equation}
\label{eqn:pot}
M^2_{1, -1, 1} \, \sim \, \left( \phi^2 + {\cal{O}}(1) \frac{\phi^4}{M_{pl}^2} \right) \, .
\end{equation}
Using \eqref{eqn:potential}, we see that the resulting potential for the modulus field $\phi$ is
(dropping the quartic term in $\phi$) 
\be
V(\phi) \, = \, e^{- 2 \Phi_d} n \sqrt{{\bf{p}}_{nc}^2 + \phi^2}
\ee
and that hence
\begin{equation}
\label{eqn:desitter}
\frac{V'}{V} \, = \, \frac{1}{\sqrt{2}} \frac{\phi}{{\bf{p}}_{nc}^2 + \phi^2}  \, 
\sim \, \frac{1}{\sqrt{2} \phi} \, , 
\end{equation}
where in the last step we have evaluated the result at late times when the momentum  ${\bf{p}}_{nc}$ in the non-compact directions is negligible (since these dimensions are expanding and the momentum hence redshifts).

Let us now make contact with the swampland criteria. First, we note from (\eqref{range}) that, since $| \theta | < \pi / 2$, the range of the modulus field does not exceed the Planck scale, in agreement with the distance conjecture \cite{Vafa1}. Next, we see from (\eqref{eqn:desitter}) that, since the field is confined to values $|\phi| < \pi / 2$, the de Sitter conjecture (\ref{cond2a}) is automatically satisfied, with a constant $c_2$ which (making use of the maximal value of $|\phi|$), is given by $c_2 = \pi / 4$.

\section{Conclusions}

We have studied shape modulus stabilization using in the context of String Gas Cosmology, and have seen that the effective potential for this modulus field is consistent with the swampland constraints (specifically, the distance and the de Sitter constraint). Note that the string gas yields a potential that stabilizes both the radial and the shape modulus fields, the radial field being stablized at the self dual radius $R = 1$ of the extra dimensions, and the shape modulus $\theta$ at the value $\theta = 0$ which corresponds to a square torus. Our analysis showed that the de Sitter conjecture is satisfied, and the value of the constant $c$ was found to be $\frac{\pi}{4}$.

Our analysis was done in the context of the simplest model for the extra dimensions, but the physics which yields the effective potentials which are consistent with the swampland constraints should be generalizable to more complicated compactifications. It is also important to point out the the origin of the potentials which we are considering are stringy. In a pure effective point particle field theory approach the winding modes which are responsible for modulus stabilization and the shape of the potentials are not present, and the effects we have discussed could not be seen.

In summary, our study provides further support for the swampland conjectures.

\section*{Acknowledgement}

GM acknowledges support from a NSERC CGSM scholarship.
RB thanks the  Pauli Center and the Institutes of Theoretical Physics and of Particle- and Astrophysics of the ETH for hospitality. The research at McGill is supported, in part, by funds from NSERC and from the Canada Research Chair program.

\end{document}